\documentclass[letterpaper, 10 pt, conference]{IEEEtran}

\usepackage[noadjust]{cite}
\usepackage{graphicx,xcolor,epstopdf}
\usepackage{amsmath,amsfonts,amssymb,amsthm,mathtools}
\usepackage{epsfig,epstopdf,url,array}
\usepackage{tikz,tikz-3dplot,tikzinclude}
\usepackage{hyperref}
\hypersetup{
	colorlinks,
	linkcolor={violet!50!black},
	citecolor={purple!50!black},
	urlcolor={blue!50!black}
}
\usepackage{enumitem}
\usetikzlibrary{intersections}
\usetikzlibrary{patterns,shapes,arrows,calc}
\usetikzlibrary{chains,positioning,quotes}
\tikzstyle{block} = [draw, rectangle, minimum height=3em, minimum width=2em]
\tikzstyle{sum} = [draw, circle,inner sep=0mm,minimum size=2mm, node distance=0.1cm, thick]
\tikzstyle{input} = [coordinate]
\tikzstyle{output} = [coordinate]
\tikzset{BlockDiagram/.style={
		block/.style = {draw, rectangle, rounded corners, text centered, minimum height=1cm, minimum width=1cm, text width=1.5cm},
		sum/.style = {draw, circle, thick, minimum size=5mm, node distance=5mm, inner sep=0mm},
		input/.style = {coordinate},
	    point/.style = {coordinate},
		output/.style = {coordinate},
}} 

\renewcommand\arraystretch{1.3}
\renewcommand{\H}{\mathcal{A}}
\renewcommand{\Re}{\operatorname{Re}}
\renewcommand{\Im}{\operatorname{Im}}

\newcommand{\MAT}[4]{\left[\begin{array}{c|c}
		#1 & #2 \\ \hline
		#3 & #4 \end{array}\right]}
\newcommand{\Q}{\mathcal{Q}}
\newcommand{\rank}[1]{\operatorname{rank}\left(#1\right)}

\newcommand{\Rset}{\mathbb{R}}

\begin{document}

\title{\textit{Robust Stability Analysis of an Uncertain Aircraft Model with Scalar Parametric Uncertainty}}

\author{\IEEEauthorblockN{Farooq Aslam\IEEEauthorrefmark{1}, Fatima Shoaib\IEEEauthorrefmark{1}, Hafiz Zeeshan Iqbal Khan\IEEEauthorrefmark{1}\IEEEauthorrefmark{2}, Muhammad Farooq Haydar\IEEEauthorrefmark{1}, and Jamshed Riaz\IEEEauthorrefmark{1}}
\IEEEauthorblockA{\IEEEauthorrefmark{1} Department of Aeronautics \& Astronautics, Institute of Space Technology, Islamabad. \\
\IEEEauthorrefmark{2} Centers of Excellence in Science and Applied Technologies, Islamabad.}}

\maketitle

\begin{abstract}
\textit{\mdseries This paper analyzes the robust stability of an uncertain aircraft model in which a key parameter, the location of the aircraft's center of gravity, is modeled as a real parametric uncertainty. A robust controller is specified and the stability bounds of the uncertain closed-loop system are determined using the small gain, circle, positive real, and Popov criteria. A graphical approach is employed in order to demonstrate the ease with which the above robustness tests can be carried out on a problem of practical interest. A significant improvement in stability bounds is observed as the analysis moves from the small gain test to the circle, positive real, and Popov tests. In particular, small gain analysis results in the most conservative robust stability bounds, while Popov analysis yields significantly less conservative bounds. This is because traditional small gain type tests allow the uncertainty to be arbitrarily time-varying, whereas Popov analysis restricts the uncertainty to be constant, real parametric uncertainty. Therefore, the results reported here indicate the conservatism associated with small gain analysis, and the effectiveness of Popov analysis, in gauging robust stability in the presence of constant, real parametric uncertainty.\\}
\end{abstract}

\begin{IEEEkeywords}
\textit{Robust Stability Analysis; Small Gain Analysis; Circle Analysis; Positive Real Analysis; Popov Analysis; Uncertain Linear Systems; Parametric Uncertainty}
\end{IEEEkeywords}

\section{Introduction}
Mathematical models of physical systems often do not cater explicitly or exactly for all of the phenomena present in those systems. Although this simplification greatly facilitates controller design, and stability and performance analysis, it also introduces unmodeled dynamic uncertainty in the mathematical abstraction. Moreover, the parameters of any real system are seldom known precisely and may even vary with time, causing parametric variations and introducing further uncertainty in the associated model. Since practical control systems must ensure stability and acceptable performance in both nominal design conditions and in uncertain or \emph{perturbed} conditions, the robustness of a control system to various parametric and unmodeled dynamic uncertainties is an important design requirement.

In classical control, gain and phase margins are typically used to measure the robustness of SISO systems to gain and phase uncertainty. However, since these margins measure only the stability/performance of the nominal closed-loop system, they are at best indirect measures of robust stability with respect to actual plant uncertainties. Fortunately, there exist several direct and reliable methods for analyzing robust stability, the most fundamental of which arise from the small gain, circle, positive real, and Popov theorems for the stability analysis of feedback systems \cite{Haddad1993b}. Important as they all are, the aforementioned methods can lead to very different assessments of robustness even when applied to the same uncertain closed-loop system. In particular, some methods can be significantly more conservative, and hence restrictive, than others.

This conservatism, or lack thereof, across methods is largely a consequence of the manner in which they characterize the underlying uncertainty \cite{Haddad1993b}. In the state space framework, the uncertainty representation is generally viewed as spanning two extremes, with the uncertainty modeled, on one hand, as a constant real parameter, and on the other, as an arbitrary time-varying real parameter. In the frequency domain, these extremes correspond to modeling the uncertainty as a transfer function either with bounded phase or with arbitrary phase, where the former corresponds to constant real parametric uncertainty, and the latter to arbitrary time-varying real parametric uncertainty. These differences in uncertainty representation have important consequences for the conservativeness of robustness estimates. For instance, the small gain, circle, and positive real criteria give robustness guarantees with respect to \emph{arbitrary} time-varying real parametric uncertainty \cite{Haddad1993}. Consequently, their application to an uncertain system with \emph{constant} real parametric uncertainty can result in very conservative robustness estimates. For situations involving constant real parametric uncertainty, the Popov criterion is more appropriate as it incorporates phase information in the frequency domain and restricts the time variation of the uncertainty \cite{Haddad1993}.

In this paper, we apply the small gain, circle, positive real, and Popov criteria to analyze the robust stability of an aircraft model in which a key parameter, the location of the aircraft's center of gravity, is modeled as a real parametric uncertainty. Although each analysis method can be applied using equivalent tests based on Lyapunov theory \cite{Haddad1993b}, the analysis here follows the graphical approach presented in \cite{Haddad1993}, wherein the authors applied the same criteria to a benchmark problem consisting of a two-mass/spring system with uncertain stiffness. Although the analysis presented there is quite insightful, the application of the above robustness tests on more complicated plants is, in general, not as straightforward as in the simple two-mass/spring case. Consequently, for the uncertain aircraft model under consideration in this paper, we adopt the more general formulation, employed in \cite{Feron1996}, for uncertain linear systems subject to real parametric uncertainties. This allows us to present the main steps involved in the analysis in a systematic manner, thereby making the approach transparent and accessible to a wider audience with fewer prerequisites, and to demonstrate the ease with which the various robustness tests can be carried out on a problem of practical interest. Moreover, in future work, we hope to use the approach presented here as a stepping stone for robust stability analysis of more complex uncertain aircraft models.

For the uncertain aircraft model under consideration, the small gain criterion results in the most conservative robust stability bounds, while Popov analysis yields significantly less conservative bounds. This highlights the effectiveness of Popov analysis when assessing robust stability in the presence of constant real parametric uncertainty. The rest of the discussion is structured as follows: Section \ref{sec2} states some fundamental results concerning robust stability analysis and uncertain linear systems; Section \ref{sec3} describes the uncertain aircraft model under consideration; Section \ref{sec4} presents the results of robust stability analysis using the small gain, circle, positive real, and Popov critera; lastly, Section \ref{sec5} concludes the discussion and suggests some directions for future work.

\section{Stability of Uncertain Linear Systems}\label{sec2}

Consider the following uncertain autonomous linear system
\begin{equation}\label{eq8}
\dot{z} = \tilde{A}z,
\end{equation}
where the state matrix is related to an uncertain real parameter $\delta\in\Rset$ as follows:
\begin{equation}\label{eq9}
\tilde{A} = A + \delta Q.
\end{equation}
In the above expressions, $z\in\Rset^n$ is the state vector, $A\in\Rset^{n\times n}$ is the nominal value of $\tilde{A}$, i.e., its value at $\delta = 0$, and $Q\in\Rset^{n\times n}$ is a fixed perturbation matrix. Moreover, the uncertain model \eqref{eq8}-\eqref{eq9} is a special case of the general formulation used in \cite{Feron1996} to express uncertain linear systems subject to real parametric uncertainties.

Suppose that $\mathrm{rank}(Q) = 1$. Then, the singular value decomposition of $Q$ is given by
\begin{equation}\nonumber
Q = \sigma vw^{\top},
\end{equation}
where $\sigma\in\Rset$ is the singular value of $Q$, and $v,w\in\Rset^n$ are the left- and right-singular vectors of $Q$. Now, consider the following single-input single-output (SISO) system
\begin{equation}\label{eq14}
\begin{split}
\dot{x} &= Ax + Bu, \\
y &= Cx.
\end{split}
\end{equation}
Suppose that its input/output matrices are given by
\begin{equation}\label{eq15}
B = \sigma v,\quad\quad C = w^{\top},
\end{equation}
and consider the feedback law
\begin{equation}\label{eq16}
u = \delta y.
\end{equation}
Then, the \emph{root locus} of the resulting closed-loop system as the uncertain parameter $\delta$ varies is identical to the locus of eigenvalues of $\tilde{A}$, as outlined below:
\begin{equation}\nonumber
\begin{split}
\dot{x} &= Ax + \sigma vu \\
&= Ax + \delta\sigma vy \\
&= Ax + \delta(\sigma vw^{\top})x = (A+\delta Q)x = \tilde{A}x.
\end{split}
\end{equation}
\begin{figure}[!ht]
	\centering
    \begin{tikzpicture}[BlockDiagram, node distance=1.0cm,>=latex']
    \node [input, name=input] {};
    \node [block, blue, right = of input, fill=blue!10,minimum height = 1cm,minimum width = 1.5cm] (P) {\Large{$M$}};
    \node [output, right = of P, node distance=3cm] (output) {};
    \node [block, violet, above of=P, node distance=1.5cm, fill=violet!10] (uncert) {\Large{$\Delta$}};

    \node [point, right of=uncert,     node distance=1.2cm] (temp1) {};
    \node [point, left  of=uncert,     node distance=1.2cm] (temp2) {};

    \draw [->,thick,violet] (P) -| node[right,pos = 0.75] {$y$} (temp1) |- (uncert);
    \draw [->,thick,violet] (uncert) -| (temp2) |- node[left,pos = 0.25] {$u$} (P);
    \end{tikzpicture}
	\caption{Standard $M$-$\Delta$ form used in robust stability analysis \cite{Skogestad2005}}\label{fig:3}
\end{figure}
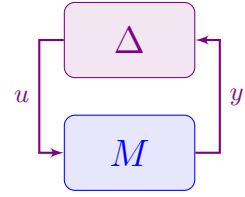
Consequently, the stability of the uncertain system \eqref{eq8} with a scalar parametric uncertainty can be analyzed using the closed-loop system \eqref{eq14}-\eqref{eq16}. Observe that the closed-loop system corresponds to the standard $M$-$\Delta$ form used in robust stability analysis \cite{Skogestad2005}. The standard form, which is depicted in Fig. \ref{fig:3}, consists of a fixed part $M$ in feedback with an uncertain part $\Delta$. For the closed-loop system \eqref{eq14}-\eqref{eq16}, the fixed part equals the transfer function $G(s)=C{(sI-A)}^{-1}B$, and the uncertain part corresponds to the uncertain parameter $\delta$. In Section \ref{sec2a}, we extend the above approach to the stability analysis of uncertain closed-loop system matrices.

Before proceeding, let us consider the rank assumption for the perturbation matrix $Q$. In general, $\mathrm{rank}(Q) = r \leq n$, and $Q = \sum_{i=1}^{r}\sigma_i u_iv_i^{\top},$ where $\sigma_i\in\Rset$ denote the singular values of $Q$, and $u_i,v_i\in\Rset^n$ its left- and right-singular vectors. If $\delta$ is a scalar parametric uncertainty, we can still express the uncertain linear system \eqref{eq8} in the $M$-$\Delta$ form for robust stability analysis. The key difference is that the fixed part of this structure, the transfer function $M$, will now be a multiple-input multiple-output system with $r$ inputs and $r$ outputs.

\subsection{Stability Analysis of Uncertain Closed-loop System}\label{sec2a}

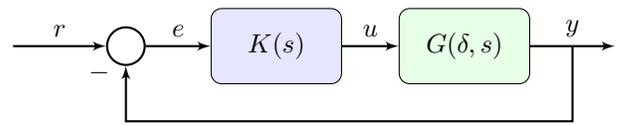
\begin{figure}[!ht]
	\centering
	\begin{tikzpicture}[BlockDiagram, node distance=1.0cm,>=latex']
	\node [input, name=input] {};
	\node [sum,   right of=input, node distance=1.5cm] (sum) {};
	\node [block, right of=sum, node distance=2cm, fill=blue!10] (K) {$K(s)$};
	\node [block, right of=K, node distance=2.5cm, fill=green!10] (G) {$G(\delta,s)$};
	\node [output, right of=G, node distance=2cm] (output) {};
	\node [input, below of=K, node distance=1cm] (temp) {};
	\draw [->,thick] (input) -- node[above,pos=0.5] {$r$} (sum);
	\draw [->,thick] (sum) -- node[above,pos=0.5] {$e$} (K);
	\draw [->,thick] (K) -- node[above,pos=0.5] {$u$} (G);
	\draw [->,thick] (G) -- node [above,pos=0.5,name=y] {$y$}(output);
	\draw [->,thick] (y) |- (temp) -| (sum);
	\node [below left of= sum,node distance = 5mm] {$-$};
	\end{tikzpicture}
	\caption{Feedback Loop with Uncertain Plant}\label{fig:1}
\end{figure}

Consider the closed-loop system depicted in Fig. \ref{fig:1} where the uncertainty $\delta$ is a real scalar. In order to study closed-loop stability in the presence of this parametric uncertainty, we would like to express the uncertain closed-loop state matrix in the $M$-$\Delta$ form discussed earlier. This section describes how the uncertainty block can be pulled out so that robust stability can be analyzed.

\renewcommand\arraystretch{1.3}

Suppose that the nominal plant and controller have minimal state-space realizations given by
\[
G(s) \equiv \MAT{A_p}{B_p}{C_p}{D_p}, \qquad K(s) \equiv \MAT{A_k}{B_k}{C_k}{D_k}.
\]
Then, the series connection of the plant and controller can be expressed as follows \cite{Bay1999}:
\begin{equation}
G(s)K(s) = \left[\begin{array}{cc|c}
A_p & B_p C_k & B_p D_k \\
0 & A_k   & B_k \\ \hline
C_p & D_pC_k  & D_pD_k \end{array}\right] \equiv \MAT{A}{B}{C}{D}. \nonumber
\end{equation}
Next, consider the nominal closed-loop system from the reference $r$ to the output $y$. For the output equation, observe that
\begin{equation}\nonumber
\begin{split}
y &= Cx + De \\
&= Cx + Dr - Dy \\
\Rightarrow y &= {(I+D)}^{-1}Cx + {(I+D)}^{-1}Dr.
\end{split}
\end{equation}
Using the output equation, the state equation can be expressed as follows:
\begin{equation}\nonumber
\begin{split}
\dot{x} &= Ax + Be \\
&= Ax + Br - By \\
\Rightarrow \dot{x} &= \big[A-B{(I+D)}^{-1}C\big]x + \big[B-B{(I+D)}^{-1}D\big]r.
\end{split}
\end{equation}
If the plant $G(s)$ is strictly proper, then $D_p = 0$, and $D=D_pD_k=0$. In that case, the nominal closed-loop state matrix is given by
\begin{equation}\label{eq11}
\begin{split}
\H &:= A - BC \\
&= \begin{bmatrix}
A_p - B_p D_k C_p & B_p C_k \\
-B_k C_p & A_k \end{bmatrix}.
\end{split}
\end{equation}
Next, consider the introduction of the scalar parametric uncertainty $\delta$ to the nominal plant matrices $A_p$ and $B_p$. In particular, consider the following perturbed matrices:
\[
\tilde{A}_p = A_p + \delta Q_A,\quad \tilde{B}_p = B_p + \delta Q_B
\]
Then, the \emph{uncertain} closed-loop state matrix is given by
\begin{eqnarray}
\tilde{\H} &:=& \begin{bmatrix}
\tilde{A}_p - \tilde{B}_p D_k C_p & \tilde{B}_p C_k \\
-B_k C_p  & A_k \end{bmatrix} \label{eq12} \\
&=& \begin{bmatrix}
A_p + \delta Q_A - (B_p + \delta Q_B) D_k C_p & (B_p + \delta Q_B) C_k \\
-B_k C_p  & A_k \end{bmatrix} \nonumber \\
&=& \H + \delta \Q, \nonumber
\end{eqnarray}
where
\begin{equation}\label{eq13}
\Q := \begin{bmatrix}
Q_A - Q_B D_k C_p & Q_B C_k  \\
0  & 0 \end{bmatrix}.
\end{equation}
In light of the discussion in the previous section, the eigenvalue analysis of the uncertain closed-loop state matrix $\tilde{\H}$, in the presence of the scalar parametric uncertainty $\delta$, can be expressed in the standard $M$-$\Delta$ form used in robust stability analysis.

\section{Aircraft Model with Parametric Uncertainty}\label{sec3}

\newcommand{\X}{\begin{bmatrix} u \\ w \\ q \\ \theta \end{bmatrix}}
\newcommand{\Xdot}{\begin{bmatrix} \dot{u} \\ \dot{w} \\ \dot{q} \\ \dot{\theta} \end{bmatrix}}
\newcommand{\U}{\eta}
\newcommand{\dm}[1]{\stackrel{\mathrm{o}}{#1}}
\newcommand{\dmh}[1]{\tilde{#1}}
\newcommand{\dd}[1]{\frac{\dm{M}_{\dot{w}}}{I_y}\left(#1\right)}
\newcommand{\ddx}[1]{\frac{\dm{X}_{\dot{w}}}{m}\left(#1\right)}
\newcommand{\dds}[1]{m_{\dot{w}}#1}
\newcommand{\ddsx}[1]{x_{\dot{w}}#1}
\newcommand{\ddsz}[1]{#1\left(1+z_{\dot{w}}\right)}
\newcommand{\Ap}{\begin{bmatrix}
\dm{X}_u & \dm{X}_w & \dm{X}_q & -mg \\
\dm{Z}_u & \dm{Z}_w & \dm{Z}_q + mV_0 & 0 \\
\dm{M}_u & \dm{M}_w & \dm{M}_q & 0 \\
0 & 0 & 1 &  0 \\
\end{bmatrix}}
\newcommand{\Bp}{\begin{bmatrix} \dm{X}_{\U} \\ \dm{Z}_{\U} \\ \dm{M}_{\U} \\ 0 \end{bmatrix}}
\newcommand{\Cp}{\begin{bmatrix}
0 & 0 & 0 & 1 \\
0 & 0 & 1 & 0 \\
0 & \frac{1}{V_0} & 0 & 0 \\
\end{bmatrix}}
\newcommand{\Dp}{\begin{bmatrix} 0 \\ 0 \\ 0 \end{bmatrix}}
\newcommand{\M}{\begin{bmatrix}
m &  -\dm{X}_{\dot{w}} & 0   & 0 \\
0 & m-\dm{Z}_{\dot{w}} & 0   & 0 \\
0 &  -\dm{M}_{\dot{w}} & I_y & 0 \\
0 &     0              & 0   & 1 \\
\end{bmatrix}}
\newcommand{\A}{\begin{bmatrix}
x_u & x_w & x_q & x_{\theta} \\
z_u & z_w & z_q & z_{\theta} \\
m_u & m_w & m_q & m_{\theta} \\
0 & 0 & 1 & 0 \\
\end{bmatrix}}
\newcommand{\B}{\begin{bmatrix} x_{\eta} \\ z_{\eta} \\ m_{\eta} \\ 0 \end{bmatrix}}
\newcommand{\Minv}{\begin{bmatrix}
\frac{1}{m} & \frac{\dm{X}_{\dot{w}}}{m(m-\dm{Z}_{\dot{w}})} & 0 & 0 \\
0 & \frac{1}{m-\dm{Z}_{\dot{w}}} & 0 & 0 \\
0 & \frac{\dm{M}_{\dot{w}}}{I_y(m-\dm{Z}_{\dot{w}})} & \frac{1}{I_y} & 0 \\
0 & 0 & 0 & 1 \\
\end{bmatrix}}
\newcommand{\Mtilde}{\begin{bmatrix}
m &  -\dm{X}_{\dot{w}} & 0   & 0 \\
0 & m-\dm{Z}_{\dot{w}} & 0   & 0 \\
0 &  -\dm{M}_{\dot{w}} + \dm{Z}_{\dot{w}}\Delta & I_y & 0 \\
0 &     0              & 0   & 1 \\
\end{bmatrix}}
\newcommand{\Mtildeinv}{\begin{bmatrix}
\frac{1}{m} & \frac{\dm{X}_{\dot{w}}}{m(m-\dm{Z}_{\dot{w}})} & 0 & 0 \\
0 & \frac{1}{m-\dm{Z}_{\dot{w}}} & 0 & 0 \\
0 & \frac{\dm{M}_{\dot{w}}-\dm{Z}_{\dot{w}}\Delta}{I_y(m-\dm{Z}_{\dot{w}})} & \frac{1}{I_y} & 0 \\
0 & 0 & 0 & 1 \\
\end{bmatrix}}
\newcommand{\QMinv}{\begin{bmatrix}
0 & 0 & 0 & 0 \\
0 & 0 & 0 & 0 \\
0 & -\frac{\dm{Z}_{\dot{w}}}{I_y(m-\dm{Z}_{\dot{w}})} & 0 & 0 \\
0 & 0 & 0 & 0 \\
\end{bmatrix}}
\newcommand{\QA}{\begin{bmatrix}
0 & 0 & 0 & 0 \\
0 & 0 & 0 & 0 \\
\mathring{Z}_u & \mathring{Z}_w & \mathring{Z}_q + \mathring{Z}_{\dot{w}}V_0 & 0 \\
0 & 0 & 0 & 0 \\
\end{bmatrix}}
\newcommand{\QB}{\begin{bmatrix} 0 \\ 0 \\ \mathring{Z}_{\eta} \\ 0 \end{bmatrix}}

In this section, we consider the linearized equations of motion that describe the decoupled longitudinal motion of an aircraft. These equations model the response of an aircraft to a disturbance acting in the aircraft's longitudinal plane of symmetry, i.e., a disturbance whose lateral component equals zero. In the following, we adopt the notation and mathematical model given in \cite{Cook2013}.

\subsection{Decoupled Longitudinal Equations of Motion}

Decoupled longitudinal motion is described by the equations for the axial force $X$, the normal force $Z$, and the pitching moment $M$. Since there is no lateral-directional motion, the lateral motion variables, their derivatives, the aerodynamic coupling derivatives, and the aerodynamic coupling control derivatives are all set to zero. Furthermore, it is assumed that the aircraft is in level flight, and that the reference axes are the same as the stability axes. With these assumptions, the equations of longitudinal symmetric motion can be stated as follows:
\begin{subequations}\label{eq1}
\begin{align}
m\dot{u} - \mathring{X}_{\dot{w}}\dot{w} =& \mathring{X}_uu + \mathring{X}_ww + \mathring{X}_qq - mg\theta + \mathring{X}_{\eta}\eta \\
(m-\mathring{Z}_{\dot{w}})\dot{w} =& \mathring{Z}_uu + \mathring{Z}_ww + (\mathring{Z}_q+mV_0)q + \mathring{Z}_{\eta}\eta \\
I_y\dot{q} - \mathring{M}_{\dot{w}}\dot{w} =& \mathring{M}_uu + \mathring{M}_ww + \mathring{M}_qq + \mathring{M}_{\eta}\eta \label{eq1c}
\end{align}
\end{subequations}
In the above equations, the state variables are given by $u$, the velocity in the $x$-direction; $w$, the velocity in the $z$-direction; $q$, the pitch-rate about the $y$-axis; and $\theta$, the attitude or pitch angle. The control input is given by the elevator deflection $\eta$. Furthermore, $m$ denotes the mass of the aircraft, $I_y$ the pitch inertia, $V_0$ the airspeed, and $g$ the acceleration due to gravity. The remaining coefficients denote various dimensional aerodynamic stability and control derivatives. For instance, $\mathring{X}_u:=\partial X/\partial u$ is the \emph{dimensional} aerodynamic stability derivative for the axial force $X$ with respect to the forward velocity $u$. The stability derivatives in \eqref{eq1} can be derived from the data for the longitudinal flight conditions and the \emph{dimensionless} derivatives, given in Tables \ref{tab:01} and \ref{tab:02}, respectively. The data is available in \cite{Cook2013}, and is for a canard-configured fly-by-wire combat aircraft.

\begin{table}
	\centering
	\caption{Longitudinal flight condition data \cite{Cook2013}}
	\label{tab:01}
	\begin{tabular}{lr}	\hline \hline
		Parameter & Value \\ 	\hline
		Altitude ($h$) & Sea level \\
        Flight path angle (at equilibrium) ($\gamma_e$) & 0 deg \\
        Angle of attack (at equilibrium) ($\alpha_e$) & 0 deg \\
        Airspeed ($V_0$) & 100 m/s \\
        Mass ($m$) & 12,500 kg \\
        Pitch inertia ($I_y$) & 105,592 $\text{kg}.{\text{m}}^2$ \\
        Air density ($\rho$) & 1.225 $\text{kg}/{\text{m}}^3$  \\
        Wing area ($S$) & 50 ${\text{m}}^2$ \\
        Mean aerodynamic chord ($c$) & 5.7 m \\
	\end{tabular}
\end{table}

\begin{table}
	\centering
	\caption{Dimensionless derivatives \cite{Cook2013}}
	\label{tab:02}
	\begin{tabular}{cccccc}	\hline \hline
		$X_u$ & 0.050 & $M_u$ & 0.003 & $Z_u$ & -1.200 \\
        $X_w$ & 0.260 & $M_w$ & 0.280 & $Z_w$ & -2.800  \\
        $X_{\dot{w}}$ & 0 & $M_{\dot{w}}$ & 0.380 & $Z_{\dot{w}}$ & -0.700 \\
        $X_q$ & 0 & $M_q$ & -0.500 & $Z_q$ & -1.200 \\
        $X_{\delta_e}$ & 0 & $M_{\delta_e}$ & 0.160 & $Z_{\delta_e}$ & -0.04 \\ \hline \hline
	\end{tabular}
\end{table}

The three equations describing longitudinal motion are coupled with an auxiliary equation relating pitch rate to attitude rate. Using the small angle assumption, this relation can be expressed as
\begin{equation}\label{eq2}
\dot{\theta} = q
\end{equation}
The outputs are taken to be the attitude $\theta$, the pitch-rate $q$, and the angle of attack $\alpha$. The last of these is related to the downward velocity $w$ and the airspeed as follows:
\[
\alpha = \frac{w}{V_0}
\]

\subsection{Nominal Aircraft Model and Controller}

The equations describing longitudinal motion \eqref{eq1}-\eqref{eq2} can be expressed compactly as
\begin{equation}\label{eq10}
M\dot{x} = A_tx + B_t\eta,
\end{equation}
where the state vector is given by
\[
x^{\top} = \begin{bmatrix} u & w & q & \theta \end{bmatrix},
\]
and $M$, $A_t$, and $B_t$ are the corresponding coefficient matrices.
Pre-multiplying both sides of \eqref{eq10} by $M^{-1}$ yields the following state space form for the equations of longitudinal symmetric motion:
\begin{equation}\label{eq3}
\dot{x} = Ax + B\eta,
\end{equation}
where the state and input matrices are given by
\begin{equation}
A = M^{-1}A_t, \quad B = M^{-1}B_t,
\end{equation}
and
\begin{equation}
M^{-1} = \Minv.
\end{equation}
Lastly, the output equation is
\begin{equation}
y = Cx + D\eta,
\end{equation}
where
\begin{equation}\label{eq4}
C = \Cp, \quad D = \Dp.
\end{equation}

The open-loop aircraft model \eqref{eq3}-\eqref{eq4} is unstable as three of its poles are located in the right-half plane. To stabilize the nominal open-loop system, we use the following feedback gains on the pitch-rate $q$ and the angle of attack $\alpha$:
\[
K_q = 1.6, \quad K_{\alpha} = 1.72.
\]
As a result, we obtain a stable \emph{augmented} system with 4 states, 3 outputs, and 1 input. The nominal aircraft model $G(s)$, specified as the transfer function from the elevator deflection $\eta$ to the attitude $\theta$ of the augmented system, is given by
\begin{equation}\label{eq17}
G(s) = \frac{2.64(s+0.0164)(s+0.635)}{(s+4.31)(s+0.68)(s^2+0.0136s+0.000327)}
\end{equation}
With reference to the feedback configuration shown in Fig. \ref{fig:1}, a nominal closed-loop system is obtained using $G(s)$ and the following controller:
\begin{equation}\label{eq5}
K(s) = \frac{3.14(s+5.14)(s+0.615)(s+0.0171)}{(s+0.356)(s+0.0175)(s^2+7.22s+13.6)}
\end{equation}
Consequently, the loop transfer function $L = GK$ has 8 states, 4 each for the plant and controller. The controller was used by the authors in an earlier work \cite{Shoaib2019} on the longitudinal control of a canard-configured high performance aircraft, and is a robust controller synthesized using $H_{\infty}$ loopshaping techniques.

\subsection{Uncertain Aircraft Model}

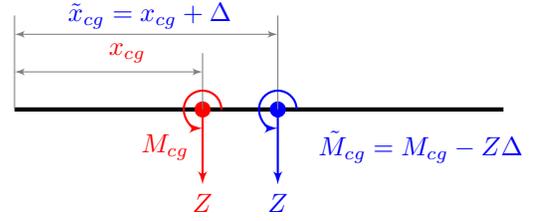
\begin{figure}[!t]
\centering
\begin{tikzpicture}[auto, node distance=1.0cm,>=latex']
	\draw [ultra thick,black] (-2.5,0) -- (4,0);
	\draw [red,fill=red] (0,0) circle [radius=0.1];
	\draw [blue,fill=blue] (1,0) circle [radius=0.1];
	
	\draw [thin,gray] (-2.5,0) -- (-2.5,1.25);
	\draw [thin,gray] (0,0) -- (0,0.75);
	\draw [thin,gray] (1,0) -- (1,1.25);
	\draw [<->,thin,gray] (-2.5,0.5) -- (0,0.5);
	\draw [<->,thin,gray] (-2.5,1.0) -- (1,1.0);
	
	\node [red] at (-1,0.75) {$x_{cg}$};
	\node [blue]at (-0.7,1.25) {$\tilde{x}_{cg} = x_{cg} + \Delta$};
	
	\node [red] at (-0.5,-0.5) {$M_{cg}$};
	\draw [->,thick,red] (0,0) -- (0,-1);
	\node [red] at (0,-1.25) {$Z$};
	\node [blue] at (2.9,-0.5) {$\tilde{M}_{cg} = M_{cg} - Z \Delta$};
	\draw [->,thick,blue] (1,0) -- (1,-1);
	\node [blue] at (1,-1.25) {$Z$};
	
	\draw [->,thick,red] (0.25,0) arc [radius=0.25, start angle=0, end angle= 270];
	\draw [->,thick,blue] (1.25,0) arc [radius=0.25, start angle=0, end angle= 270];
	\end{tikzpicture}
\caption{Variation in $x_{cg}$}\label{fig:2}
\end{figure}

Next, we describe the parametric uncertainty resulting from imprecise knowledge of $x_{cg}$, the location of the aircraft's center of gravity. Suppose that the actual \emph{cg} is located at
\[
\tilde{x}_{cg} = x_{cg} + \Delta,
\]
where $\Delta$ is the associated absolute uncertainty. Then, the pitching moment about the actal $cg$ varies as shown in Fig. \ref{fig:2}. This variation in turn introduces perturbations in the dimensional derivatives used in the pitching moment equation Eq. \eqref{eq1c}. In particular, we obtain the following uncertain dimensional derivatives:
\begin{eqnarray}
{\dmh{M}}_{\dot{w}} &=& \mathring{M}_{\dot{w}} - \mathring{Z}_{\dot{w}} \Delta \nonumber \\
{\dmh{M}}_u &=& \mathring{M}_u - \mathring{Z}_u \Delta \nonumber \\
{\dmh{M}}_w &=& \mathring{M}_w - \mathring{Z}_w \Delta \nonumber \\
{\dmh{M}}_q &=& \mathring{M}_q - \mathring{Z}_q \Delta \nonumber \\
{\dmh{M}}_{\eta} &=& \mathring{M}_{\eta} - \mathring{Z}_{\eta} \Delta \nonumber
\end{eqnarray}

Replacing the nominal dimensional derivatives ${\mathring{M}}_{(\cdot)}$ by their uncertain counterparts ${\dmh{M}}_{(\cdot)}$, for each subscript $\{u,w,\dot{w},q,\eta\}$, we obtain the \emph{perturbed} dynamical system
\[
\tilde{M}\dot{x} = \tilde{A}_px + \tilde{B}_p\eta,
\]
where $\tilde{M}$, $\tilde{A}_p$, and $\tilde{B}_p$ are the uncertain versions of $M$, $A_p$, and $B_p$, respectively. Pre-multiplying by ${\tilde{M}}^{-1}$ yields the following perturbed state space system
\[
\dot{x} = \tilde{A} + \tilde{B}\eta,
\]
where the state and input matrices are given by $\tilde{A} = {\tilde{M}}^{-1}\tilde{A}_p$, and $\tilde{B} = {\tilde{M}}^{-1}\tilde{B}_p$, respectively. In particular,
\[
{\tilde{M}}^{-1} = M^{-1} + \Delta\QMinv.
\]
Furthermore, using simple algebraic manipulation, we observe that
\begin{equation}\nonumber
\begin{split}
\tilde{A} &= A + \Delta Q_A,\\
\tilde{B} &= B + \Delta Q_B,
\end{split}
\end{equation}
where
\[
Q_A = -\mu\QA,\quad Q_B = -\mu\QB,
\]
and $\mu = m/\big[I_y(m-\mathring{Z}_{\dot{w}})\big]$. This shows that $\rank{Q_A}=\rank{Q_B}=1$. Lastly, substituting for $Q_A$, $Q_B$, $C_p$, $C_k$, and $D_k$ in \eqref{eq13}, we observe that $\mathrm{rank}(\Q) = 1$.

\section{Robust Stability Analysis of Uncertain Aircraft Model}\label{sec4}

In this section, we apply different methods for robust stability analysis to the uncertain closed-loop system formed using the perturbed aircraft model and the nominal controller \eqref{eq5}. In particular, we use the small gain, circle, positive real, and Popov criteria to determine the range of values of the parametric uncertainty $\Delta$ for which the uncertain closed-loop is stable. To this end, our focus is on the uncertain closed-loop state matrix $\tilde{\H}$, given in \eqref{eq12}.

In order to characterize the differences in the results obtained using the aformentioned analysis tools, we first determine the exact range for $\Delta$ such that the uncertain closed-loop is stable. Since the perturbation matrix $\Q$ has rank 1, it follows from the discussion in Section \ref{sec2} that in the $M$-$\Delta$ form corresponding to the uncertain closed-loop state matrix, the transfer function $M$ is SISO. Consequently, for this example, the stability of the uncertain closed-loop system can be guaged using classical tools such as root locus and gain margin. Here, we use the latter to determine the exact stability range for which the uncertain closed-loop system is stable. In particular, we find that the exact stability range is given by
\begin{equation}\label{eq6}
-16.3548 \leq \Delta \leq 0.512838.
\end{equation}
With reference to Figure \ref{fig:2}, if the \emph{cg} is located $0.513\text{m}$ to the right of its nominal value, the closed-loop will become unstable. The discrepancy between the lower and upper bounds is due to the fact that a rearward shift of the \emph{cg}, in general, reduces aircraft longitudinal stability, and a forward shift improves longitudinal stability \cite{Nelson1989}. Next, following the same graphical approach as employed by the authors in \cite{Haddad1993}, we analyse robust stability using the small gain, circle, positive real, and Popov criteria. In particular, for each method, we determine the corresponding stability range for $\Delta$, and compare the result with the exact stability range \eqref{eq6}.

\subsection{Small Gain Analysis}

\begin{figure}
	\centering
    \includegraphics[width=0.94\linewidth]{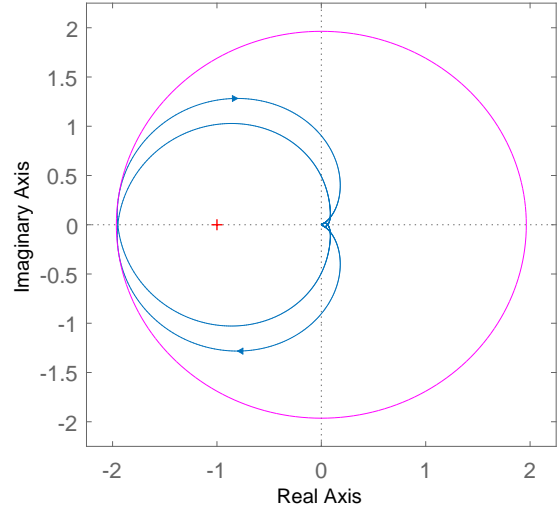}
	\caption{Robust stability analysis using the small gain criterion}
	\label{fig:smallgain}
\end{figure}

In order to analyze robust stability using the small gain criterion, we begin by drawing the Nyquist plot of $M$, i.e., we plot $\Im[M(j\omega)]$ vs. $\Re[M(j\omega)]$. Next, we find the smallest circle, centered at the origin of the complex plane, which completely encompasses the Nyquist plot without intersecting with it. Robust stability analysis then amounts to determining the points of intersection of this circle with the positive and negative real-axes. In particular, the positive real-intercept equals $-1/\underline{\Delta}$, and the negative real-intercept equals $-1/\overline{\Delta}$. Therefore, the stability range can be determined as follows:
\[
\underline{\Delta} = -\frac{1}{r_{\text{sg}}},\quad\quad\overline{\Delta} = \frac{1}{r_{\text{sg}}},
\]
where $r_{\text{sg}}$ is the radius of the smallest circle satisfying the small gain criterion. For the uncertain aircraft model under consideration, the Nyquist plot of $M(s)$ and the smallest circle that encompasses it are shown in Fig. \ref{fig:smallgain}. The circle has a radius of 1.9639. Consequently, the stability bounds predicted by small gain analysis are given by
\begin{equation}\label{eq7}
-0.5092 \leq\Delta\leq 0.5092.
\end{equation}
Compared to the exact stability bounds in \eqref{eq6}, we see that the lower limit is quite conservative.

\subsection{Circle Analysis}

\begin{figure}
	\centering
    \includegraphics[width=0.94\linewidth]{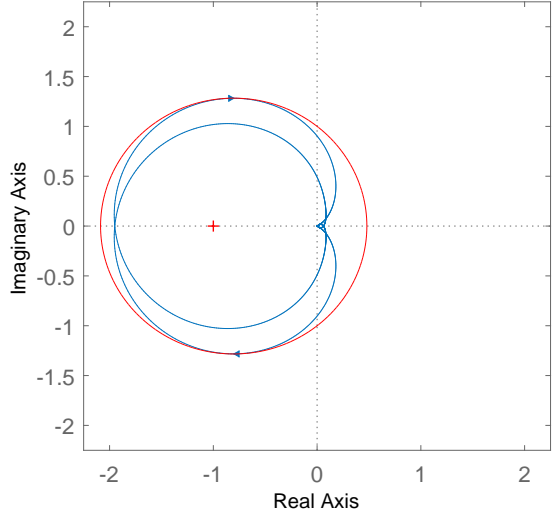}	
	\caption{Robust stability analysis using the circle criterion}
	\label{fig:circle}
\end{figure}

The circle criterion is similar to the small gain criterion in that here also we find a circle that wholly encompasses the Nyquist plot of $M$. However, we can now center the circle at any point on the real axis. Once such a point is selected and a sufficiently large circle is drawn, the stability bounds can be inferred from its real-axis intercepts. Just as in small gain analysis, the circle's positive real-intercept equals $-1/\underline{\Delta}$, and its negative real-intercept equals $-1/\overline{\Delta}$.

In particular, suppose that $x_{\text{max}}$ and $x_{\text{min}}$ denote, respectively, the maximum and minimum values of the real part of the Nyquist diagram of $M$, and consider a circle centered at $x_c = (x_{\text{max}}+x_{\text{min}})/{2}$. Then, the stability bounds can be determined as follows:
\[
\underline{\Delta} = -\frac{1}{x_c+r_c},\quad\quad\overline{\Delta} = -\frac{1}{x_c-r_c},
\]
where $r_c$ denotes the radius of the smallest circle satisfying the circle criterion.

Figure \ref{fig:circle} plots the Nyquist diagram for the uncertain aircraft model, and a circle that satisfies the circle criterion centered at $x_c = -0.8036$ with radius $r_c = 1.2834$. Consequently, the stability bounds predicted by circle analysis are given by
\begin{equation}
-2.0845 \leq\Delta\leq 0.4792.
\end{equation}
Compared to the stability bounds \eqref{eq7} given by small gain analysis, we see that circle analysis reduces the conservatism for one bound but increases it for the other. A similar result was reported in \cite{Haddad1993}.

\subsection{Positive Real Analysis}

\begin{figure}
	\centering
    \includegraphics[width=0.94\linewidth]{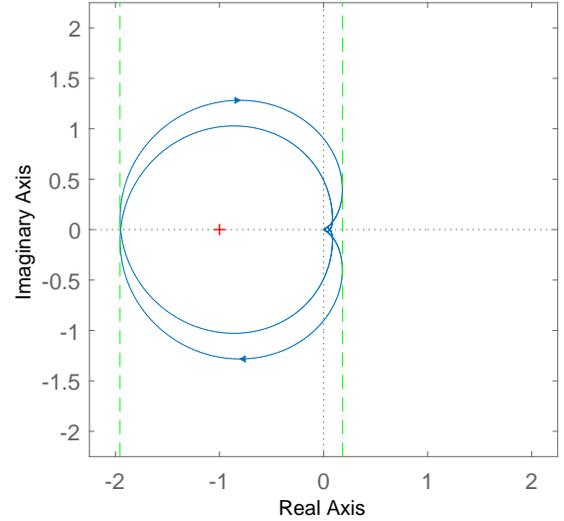}
	\caption{Robust stability analysis using the positive real criterion}
	\label{fig:positivereal}
\end{figure}

The positive real criterion extends the circle criterion by considering circles whose centers tend towards infinity along the positive and negative real axes. For instance, suppose a circle is centered at a point which lies far to the right of the origin. Then, one can draw a circle which intersects the negative real axis just to the left of $x_{\text{min}}$, the minimum value of the real part of the Nyquist plot of $M$, thereby encompassing the Nyquist diagram. This is equivalent to drawing a vertical line passing through $x_{\text{min}}$. Placing the center of the circle far to the left of the origin and repeating the above steps yields a positive real-axis intercept passing through $x_{\text{max}}$, the maximum value of the real part of the Nyquist plot. As before, the negative real-axis intercept equals $-1/\overline{\Delta}$, and the positive real-axis intercept equals $-1/\underline{\Delta}$. Thus, the stability bounds are given by
\[
\underline{\Delta} = -\frac{1}{x_{\text{max}}},\quad\quad\overline{\Delta} = -\frac{1}{x_{\text{min}}},
\]

Figure \ref{fig:positivereal} plots the Nyquist diagram and the positive real criterion for the uncertain aircraft model. The stability bounds predicted by positive real analysis are given by
\begin{equation}
-5.4866 \leq\Delta\leq 0.5112.
\end{equation}
Compared to the stability bounds calculated earlier, we see that positive real analysis reduces conservatism in both the upper and lower bounds.

\subsection{Popov Analysis}

\begin{figure}
	\centering
    \includegraphics[width=0.94\linewidth]{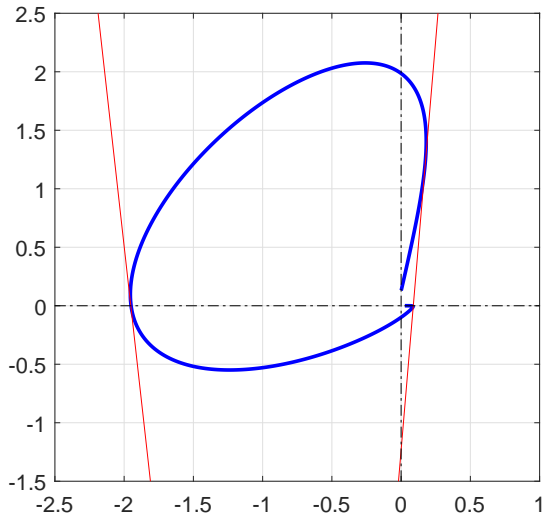}
	\caption{Robust stability analysis using the Popov criterion}
	\label{fig:popov}
\end{figure}

Popov analysis is based on a modified Nyquist diagram of $M$. In particular, we plot $\Im[\omega M(j\omega)]$ vs. $\Re[M(j\omega)]$ for $\omega\geq0$. Then, we find straight lines which intersect the positive and negative real-axes such that they are as close to the origin as possible and lie to the right and left, respectively, of the Popov diagram, i.e., the lines do not intersect the Popov diagram. These lines are known as the \emph{Popov lines}, their slopes denote \emph{Popov multipliers}, and their real-axes intercepts are related to the stability bounds. As before, the positive real-intercept equals $-1/\underline{\Delta}$, and the negative real-intercept equals $-1/\overline{\Delta}$. Popov analysis is closely related to positive real analysis. In particular, choosing the lines to be vertical gives the same stability bounds as those for positive real analysis \cite{Haddad1993}.

Figure \ref{fig:popov} plots the Popov diagram and the Popov lines for the uncertain aircraft model. The stability bounds predicted by Popov analysis are given by
\begin{equation}
-11.3692 \leq\Delta\leq 0.5123.
\end{equation}
Compared to the stability bounds calculated earlier, we see that Popov analysis gives the least conservative stability bounds, particularly with respect to the lower stability bound. For this bound, a significant improvement is seen as we move from the small gain criterion, to the circle, positive real, and Popov criteria. The improvement in stability bounds, summarized in Table \ref{tab:03}, is in accordance with the discussion in the Introduction, as well as with earlier results reported in literature \cite{Haddad1993}.

\begin{table}[bp]
	\centering
	\caption{Stability bounds for different analysis methods}
	\label{tab:03}
	\begin{tabular}{lr}	\hline \hline
		Analysis & Stability bounds \\ 	\hline
        Exact & $-16.3548 \leq \Delta \leq 0.5128$ \\
		Small gain & $-0.5092 \leq\Delta\leq 0.5092$ \\
        Circle & $-2.0845 \leq\Delta\leq 0.4792$ \\
        Positive real & $-5.4866 \leq\Delta\leq 0.5112$ \\
        Popov & $-11.3692 \leq\Delta\leq 0.5123$ \\ \hline \hline
	\end{tabular}
\end{table}

\section{Conclusion}\label{sec5}

In this paper, the robust stability of an uncertain aircraft model was analyzed using the small gain, circle, positive real, and Popov criteria. The plant uncertainty is a scalar parametric uncertainty, namely uncertainty associated with the location of the aircraft's center of gravity. A robust controller was designed for the nominal system, and the robustness of the uncertain closed-loop was analyzed. For each method, the analysis was carried out graphically in order to illustrate the ease with which the methods can be applied to a problem of practical interest. A significant improvement in stability estimates was observed as the analysis moved from the small gain test to the circle, positive real, and Popov tests. In particular, the Popov test yielded the least conservative bounds. The differences in conservatism across the four methods are attributable primarily to how each method characterizes the underlying uncertainty. In particular, small gain type tests allow the uncertainty to be arbitrarily time-varying, whereas the Popov test restricts the uncertainty to be constant, real parametric uncertainty \cite{Haddad1993}. In the frequency domain, this amounts to modeling the uncertainty as a transfer function either with arbitrary phase (small gain type tests) or with bounded phase (Popov test).

Some directions for future work include extending the approach presented in this paper to other scenarios involving uncertain linear systems with parametric and unmodeled dynamic uncertainties. A natural extension is to consider situations with scalar parametric uncertainties where the perturbation matrix has rank greater than $1$. Since the fixed part of the standard $M$-$\Delta$ structure will no longer be SISO, relaxing the rank assumption will entail using the MIMO counterparts of the Nyquist, small gain, circle, positive real, and Popov criteria. Another possibility is to explore equivalent Lyapunov function-based tests for uncertain linear systems with multiple constant real parametric uncertainties, such as those developed in \cite{Feron1996}. Given that analysis methods often form the basis for associated synthesis methods, it might also be interesting to compare robust controllers synthesized using techniques based on different analysis criteria.

\bibliography{References}
\end{document}